\pdfoutput=1

\documentclass[10pt,twocolumn,reprint,amsmath,amssymb,amsfonts,aps,prd,superscriptaddress,showpacs,nofootinbib]{revtex4-2}
\usepackage{graphicx}
\usepackage{hyperref}
\usepackage[normalem]{ulem}
\usepackage{slashed}
\usepackage{float}
\usepackage[caption=false]{subfig}
\usepackage{bm}

\hypersetup{
	bookmarks=true,         
	unicode=true,          
	pdftoolbar=true,        
	pdfmenubar=true,        
	pdffitwindow=false,     
	pdfstartview={FitH},    
	pdftitle={ResonantADM},    
	pdfauthor={Author},     
	pdfsubject={Subject},   
	pdfcreator={Creator},   
	pdfproducer={Producer}, 
	pdfkeywords={keyword1} {key2} {key3}, 
	pdfnewwindow=true,      
	colorlinks=true,       
	linkcolor=black,          
	citecolor=black,        
	filecolor=black,      
	urlcolor=black           
}

\newcommand{\be}{\begin{equation}}
\newcommand{\ee}{\end{equation}}
\def\bsp#1\esp{\begin{split}#1\end{split}}

\newcommand{\corsika}{\texttt{CORSIKA}}

\def\aem{\alpha_{\rm em}}

\def\tl{\mathcal{T}}

\begin{document}
	
\title{Atmospheric resonant production for light dark sectors}

\newcommand*{\IPII}{Institut de Physique des 2 Infinis de Lyon (IP2I),
UMR5822, CNRS/IN2P3, F-69622 Villeurbanne Cedex, France}

\author{\vspace{0.5cm} L.~Darm\'e}
\affiliation{\IPII}

	\begin{abstract}
	Cosmic ray atmospheric showers provide an effective environment for the production of MeV-scale dark sector particles.
	We show that, when available, the resonant annihilation of positrons from the shower on atmospheric electrons is the dominant production mechanism by more than an order of magnitude. We provide a quantitative example based on dark photon production and update existing constraints on a corresponding light dark matter model from kilotons neutrino experiments and xenon-based direct detection experiments.

	\end{abstract}
		\vspace{1cm}
		\maketitle
	\setcounter{footnote}{0}

\section{Introduction}

The idea of using the abundant cosmic ray (CR) flux impinging on earth's atmosphere as an ``atmospheric collider'' has a rich  history dating back to the discovery of muons. In recent years, this flux has been widely leveraged to search for Feebly interacting particles (FIPs) with mass ranging from the MeV to the GeV scale whose interaction with matter is suppressed enough that they may have escaped prior detection~\cite{Alexander:2016aln,Battaglieri:2017aum,Beacham:2019nyx,Lanfranchi:2020crw}. The case where this FIP acts as a mediator between the Standard Model (SM) particles and a light, sub-GeV, dark matter candidate is particularly compelling. Indeed, while such construction preserves most of the elegant features of the vanilla ``WIMP'' dark matter, the small dark matter mass coupled with the low velocity of the galactic dark matter halo dramatically weakens existing direct detection searches. If this light dark matter (LDM) is assumed to compose all or a large fraction of the total dark matter density, strong constrains can nonetheless be obtained by estimating the CR scattering on dark matter particles. This leads to a secondary, relativistic flux which can eventually leave a sizeable recoil signature, either in direct detection experiments~\cite{An:2017ojc,Emken:2017hnp,Ema:2018bih,Bringmann:2018cvk,Cappiello:2018hsu,Krnjaic:2019dzc,Bondarenko:2019vrb,Emken:2021vmf,Xia:2021vbz,An:2021qdl} or in neutrino experiments~\cite{Cappiello:2019qsw,Guo:2020drq,Granelli:2022ysi,Wang:2021jic}. 
Alternatively, one can specified a proper model for a stable light particle which may not constitute the actual relic dark matter or only a small fraction. Direct production of such light states from CR interactions in the atmosphere is however possible, with a subsequent detection via scattering signatures in direct detection or neutrino detectors~\cite{Alvey:2019zaa,Plestid:2020kdm,Harnik:2020ugb,Kachelriess:2021man,Arguelles:2022fqq,Chauhan:2021fzu}.

In this work, we build on the second approach by pointing out a new production mechanism relying on the annihilation of CR shower-induced positrons on atmospheric electrons. We illustrate this mechanism for models of LDM based on the exchanged of a new massive vector mediator. For definiteness, we mostly focus on a massive vector sharing the same interaction pattern as a photon (thus called dark photon) and a dark matter model following an ``inelastic dark matter" structure as advocated in~\cite{Izaguirre:2015zva,Izaguirre:2017bqb}. Our result could however equally apply to scalar or Majorana dark matter up to order one factors. 
We will further translate this improved description of the production rates by estimating the limits and projections from electron scattering in SuperK~\cite{Super-Kamiokande:2011lwo} and HyperK~\cite{Hyper-Kamiokande:2018ofw}, and from coherent nuclear scattering in xenon-based dark matter experiments~\cite{XENON:2018voc,XENON:2020kmp,DARWIN:2016hyl}.

\section{Dark sector fluxes from cosmic rays}
\label{sec:LDMprod}

\subsection{Cosmic ray showers description} 

Cosmic ray showers  originate primarily from high energy protons impinging on the high atmosphere. The resulting showers are therefore hadronic at first, with electromagnetic components appearing afterwards mostly as byproducts of the decay of $\pi^0$ mesons.
We will focus on dark sector states that are either stable, or with a decay length significantly longer than the typical length of the shower. Since the cosmic ray flux is nearly isotropic, the longitudinal and transverse development of the shower can then be integrated out in favour of an effective energy-dependent ``track-length flux''~\cite{Celentano:2020vtu,Darme:2021sqm}:
\begin{align}
\label{eq:tlanalytical}
\tl_\pm (E) ~=~   \int_0^\infty d E_0 \frac{d \Phi^{\rm{CR}}}{d E_0} \left( \int_0^\infty d\ell ~ n_e (\ell) \frac{\rho(\ell)}{\rho_0} \right)\ ,
\end{align}
where $\frac{d \Phi^{\rm{CR}}}{d E_0}$ is the differential CR flux 
, $n_e$ is the differential number density of $e^\pm$ in the shower, function of the shower depth parameter $\ell$ which parametrises the longitudinal shower development, and $\rho$ (resp. $\rho_0$) is the atmospheric density at shower depth $\ell$ (resp. at ground level).\footnote{We use the parameters for the U.S. standard atmosphere as implemented in \corsika~\cite{Heck:1998vt}.} The ratio in densities is used to normalised the track length flux to a ground level ``target'' atmosphere. 

This quantity can be calculated once and for all and then used to estimate the dark sector differential fluxes $\frac{d \Phi}{dE_\chi}$ for a variety of new physics models following
\begin{align}
\frac{d \Phi}{dE_\chi}= \frac{\mathcal{N}_A \rho_0}{A} \int_0^{\infty}  dE~ \tl_\pm (E)~  \frac{d\sigma}{dE_\chi} \ ,
\end{align}
where $\frac{d\sigma}{dE_\chi}$ is the relevant production differential cross-section from the interaction between the shower particles and the atmosphere (it hence typically includes a factor of the atomic number $Z$ or $Z^2$ depending on the relevant process).

In practice, we obtain this track length flux from two different approaches. First a semi-analytical way, based on the method advocated in~\cite{Darme:2021sqm}. The differential number density of secondary neutral mesons $n_{M_0}(E)$ from a $pN$ collision is obtained in this case from the \texttt{QGSPJETII} software~\cite{Ostapchenko:2010vb} at energy below $10$ GeV and by \texttt{EPOS-LHC}~\cite{Pierog:2013ria} above, as packaged in \texttt{CRMC}~\cite{ulrich_ralf_2021_4558706}. This energy-dependent spectrum is then convoluted with the cosmic ray energy flux as parametrised in~\cite{Boschini:2017fxq}. The subsequent electromagnetic showers are then described analytically following the approach of Rossi and Griesen~\cite{RevModPhys.13.240} as reviewed and expanded in~\cite{Lipari:2008td}. Second, we use a purely numerical approach by relying on the software \corsika~\cite{Heck:1998vt} to simulate  cosmic ray showers for incoming protons with kinetic energy between  $2.5$ GeV and $15$ TeV (using \texttt{EPOS} for the hadronic interaction). The track length is obtained directly from the simulated tracks of photons and $e^\pm$, and the meson differential energy distributions collected for both the $\pi^0$ and $\eta$ mesons.\footnote{In practice we sum the path lengths for each track, obtained from the travelled length and the mass overburden at its start and end point, see~\cite{Heck:1998vt}.}
We average over azimuthal angles for the shower proton progenitors.

\begin{figure}
	\centering
		\includegraphics[width=0.45\textwidth]{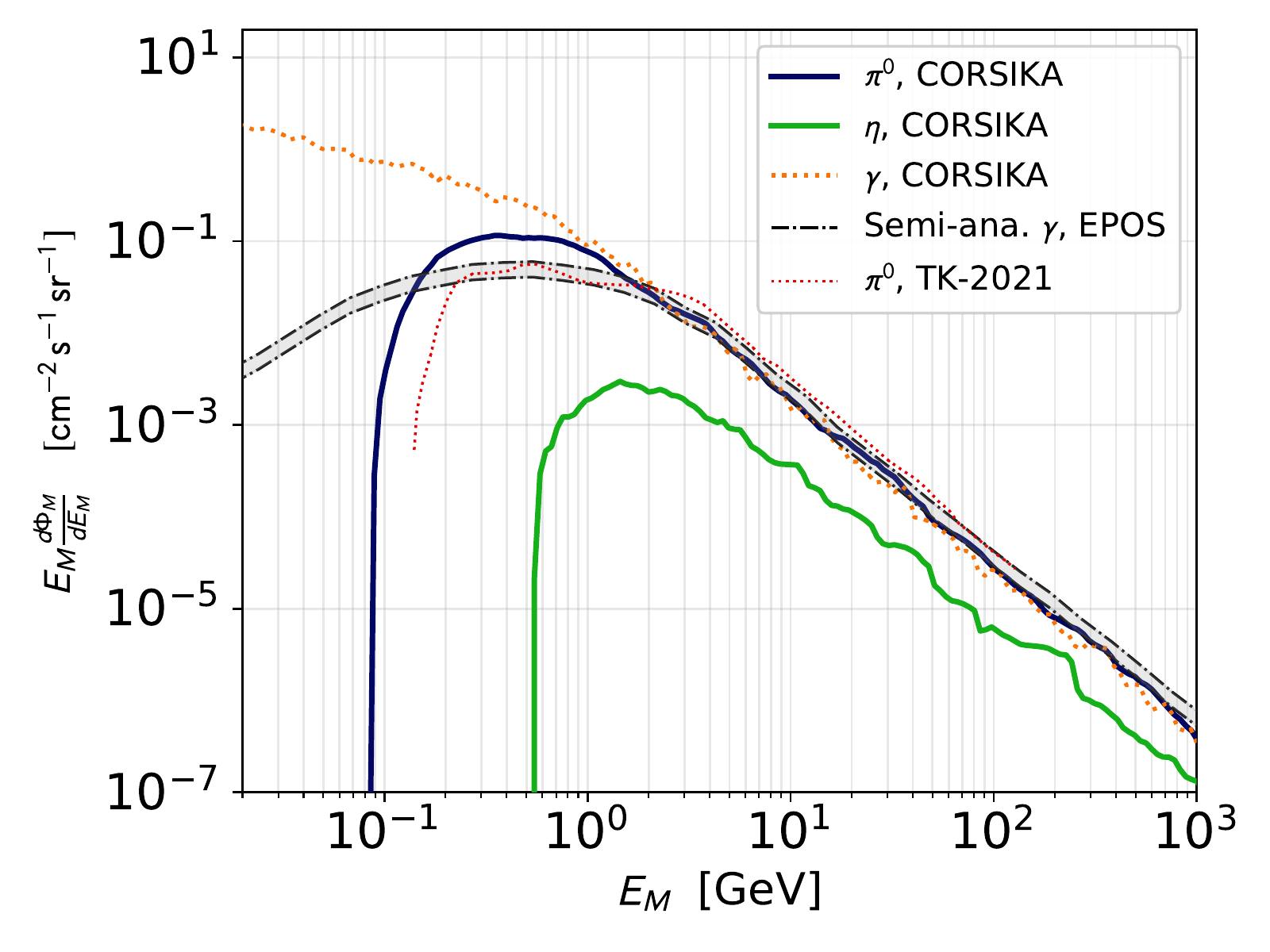}
	\caption{Differential $\pi^0$ and $\eta$ mesons fluxes as function of the meson energy. The blue and green line are the result from the \corsika \ simulation, the grey area is the result of the semi-analytical procedure. The red dotted line is the $\pi^0$ flux found in~\cite{Kachelriess:2021man}.}
			\label{fig:mesflux}
\end{figure}

\begin{figure}
	\centering
		\includegraphics[width=0.45\textwidth]{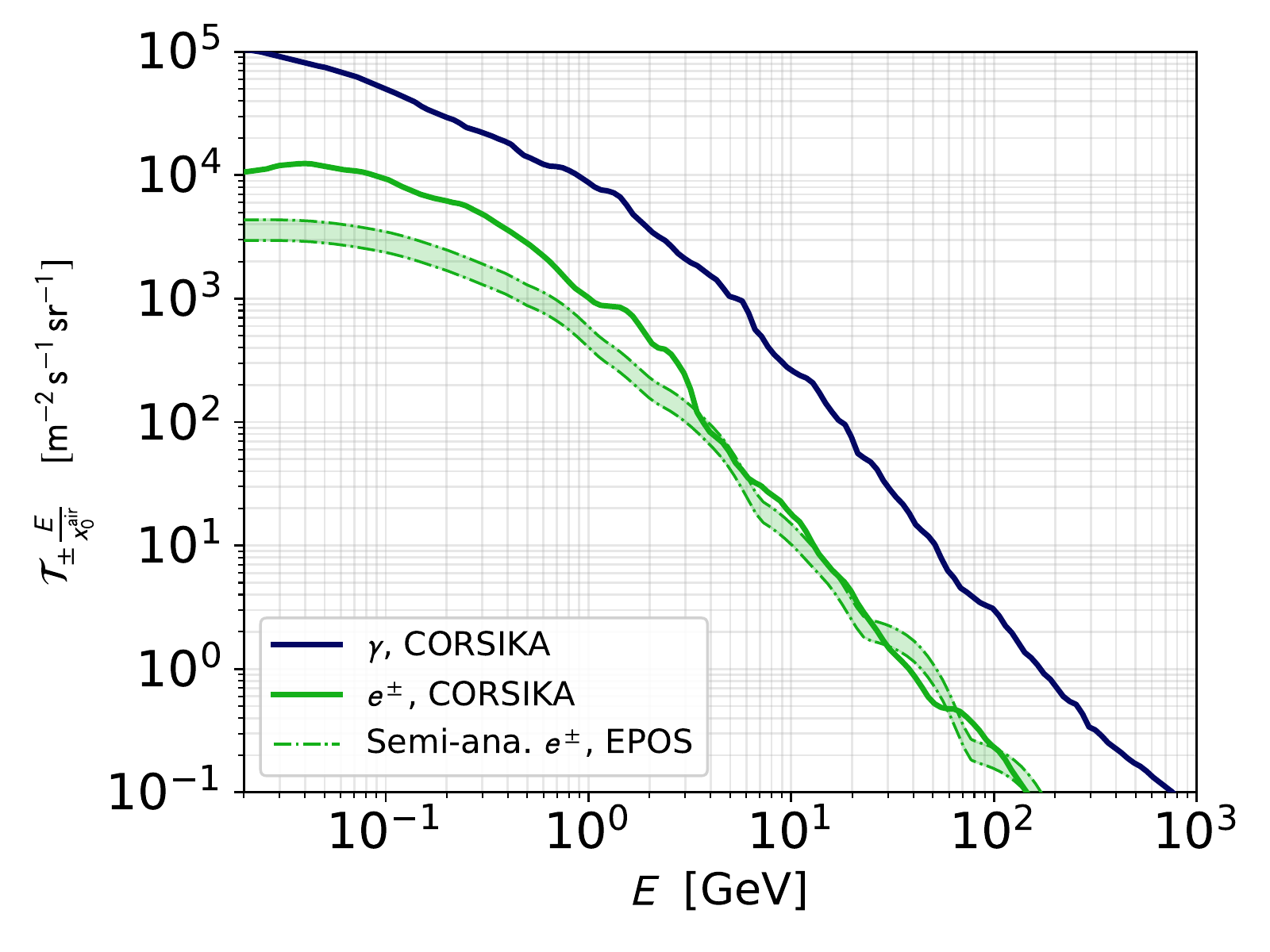}
	\caption{Differential track length fluxes in unit of atmospheric radiation length at ground level, $x_0^{\rm{air}} = 30.4$m  for both $e^-,e^+$ (green line) and for $\gamma$ (blue line) as function of their energy. The light green region corresponds to the result from the semi-analytical procedure.}
			\label{fig:TLE}
\end{figure}

We use the first approach for cross-checking purposes and rely on the fully numerical second approach in our final limits and projections. 
We present the resulting meson distribution in Fig.~\ref{fig:mesflux} and the track length fluxes in Fig.~\ref{fig:TLE}, with the corresponding datasets available at a \href{https://zenodo.org/record/6561236}{Zenodo repository}.\footnote{https://zenodo.org/record/6561236} The two approaches present a good agreement, with as expected larger fluxes at low energies for the fully numerical approach. This reflects the fact that it further includes the secondary mesons neglected in the semi-analytical method. We additionally make a rough estimate of the error of the semi-analytical procedure by using either the nuclear collision length or the nuclear interaction length when estimating the distance travelled by CR proton before its energy is significantly reduced (see~\cite{Celentano:2020vtu}). Our results are in good agreement with the recent literature for mesons productions in the atmosphere~\cite{ArguellesDelgado:2021lek,Kachelriess:2021man}. Note that the production rates have a $\sim 50 \%$ theoretical uncertainty due to the dependence on the hadronic interaction models used to model mesons production in the shower~\cite{ArguellesDelgado:2021lek}.

\subsection{Dark sector productions} 

We will focus for definiteness on simple LDM scenarios with a dark photon mediator $V^\mu$ (with mass $m_V$) interacting with the SM electromagnetic current $\mathcal{J}_{\rm{em}}^{\mu}$ via a small kinetic mixing parameter $\varepsilon$ and with a dark sector current $ \mathcal{J}_{D}^{\mu}$ with a large gauge coupling $g_D ~\equiv~ \sqrt{4 \pi \alpha_D}$:
\begin{align}
     \mathcal{L} \supset - V_\mu \left( e \varepsilon \mathcal{J}_{\rm{em}}^{\mu} + g_D \mathcal{J}_{D}^{\mu} \right)\ .
\end{align}
Since direct on-shell production of the dark photon dominates the production rates, the precise nature of the dark matter does not impact significantly the result of this work. In order to compare with existing limits, we consider the case of an inelastic dark matter structure:
\begin{equation}
\mathcal{J}_{D}^{\mu} = - i \, \, \overline \chi_2 \gamma^\mu \chi_1 \ ,
\end{equation}
with very small splitting between both states $m_{\chi_1} \sim m_{\chi_2} ~\equiv~ m_\chi$. Assuming $m_V \lesssim 2 m_\chi$, the dark photon decays mostly to dark matter states.

The first source of dark photons is the decay of neutral mesons, $\pi^0 \rightarrow \gamma V$, $\eta, \eta' \rightarrow \gamma V$. The typical branching ratio is given by
\begin{align}
\mathcal{BR}(\pi^0\rightarrow V \gamma) = 2 \varepsilon^2 \left( 1- \frac{m_V^2}{M_{\pi^0}^2}\right)^3 \times \textrm{BR} (\pi^0 \to \gamma \gamma) \ ,
\end{align}
and similarly for $\eta \to \gamma V$. Thanks to the long life-time of these mesons, this process is only  suppressed by two powers of the small kinetic mixing parameter.

On the other hand, cosmic ray showers develop a large electromagnetic component from the radiative decays of those same light neutral mesons $\pi^0, \eta$. These showers convert the large initial energy of the primary meson into a high number of low energy electrons and positrons. The latter are particularly interesting as they can annihilate with the electrons present in the air with a cross-section given by:
\begin{align}
\sigma_{\rm res} = \frac{2 \pi^2\varepsilon^2 \aem }{m_e} \delta ( E_+-\frac{m_V^2}{2m_e})  \equiv \tilde{\sigma}_{\rm res} \, \delta ( E_+-E_{\rm res}) \ .
\end{align}

\begin{figure}
	\centering
		\includegraphics[width=0.45\textwidth]{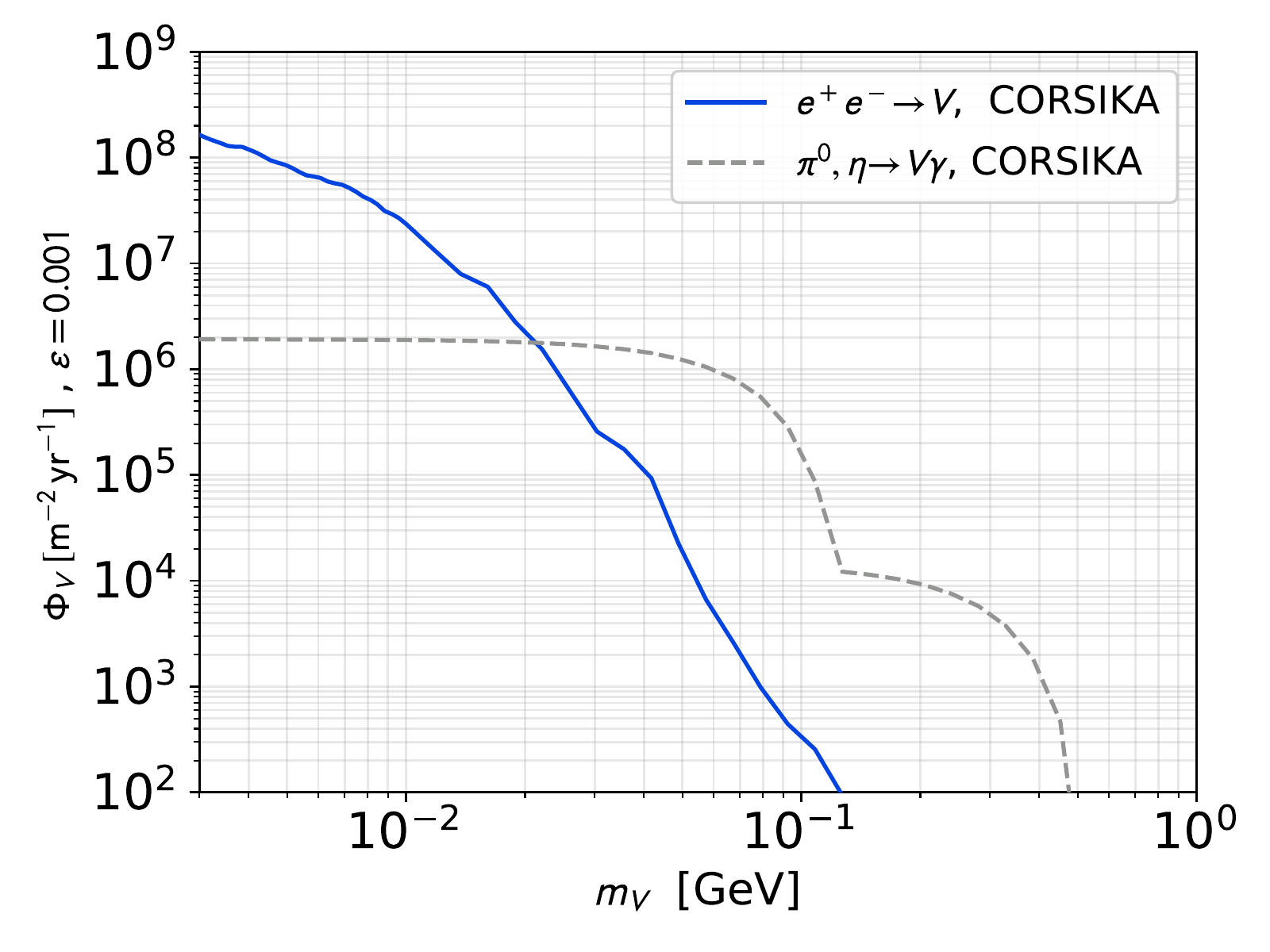}
	\caption{Dark photon flux as function of its mass $m_V$ at $\varepsilon=0.001$ from both $\pi^0$ and $\eta$ mesons decay (dashed grey line) and from the resonant annihilation of CR shower $e^+$ (thick blue line). }
			\label{fig:FluxesV}
\end{figure}

We show in Fig.~\ref{fig:FluxesV} the resulting dark photon fluxes for both the resonant and meson decay production channels for two typical dark matter masses. At low mass the resonant production dominates the $\pi^0,\eta \to V\gamma$ process by more than an order of magnitude, and generate a large flux of ``monochromatic'' dark photons with energies $m_V^2/2m_e$.\footnote{In the limit of a very light mediator, one recovers the case of millicharge particles. In particular, the relevant processes become $e^+ e^- \to \chi \chi$ and $\pi^0 \to \gamma  \chi \chi$. We expect a similar enhancement of the production fluxes for MeV-scale millicharge particles and leave to future work a thorough study of this scenario.}

\subsection{LDM detection}

We consider first the $\chi - e$ scattering process. The differential cross section for $\chi e \rightarrow \chi e$ scattering with respect to the outgoing electron energy $E_f$ in the laboratory frame is~\cite{Batell:2014mga}:  
\begin{align}
\label{eq:csscat}
\frac{d \sigma_{f,s}}{d E_f} =  4 \pi \varepsilon^2 \aem \alpha_D \frac{2 m_e E^2 - f_{f,s}(E_f)(E_f - m_e)}{(E^2 - m_\chi^2)(m_{V}^2 + 2 m_e E_f - 2 m_e^2)^2} 
\end{align}
where $E$ is the incoming LDM energy and $f$ and $s$ stand for the Dirac fermion and scalar $\chi$ respectively; $f_f(E_f) = 2 m_e E -m_e E_f + m_\chi^2 + 2 m_e^2$, $f_s(E_f) = 2 m_e E + m_\chi^2$. The total signal yield can then be obtained analytically, convoluting the differential cross section with the incoming LDM distribution and the cut efficiency for electron recoil detection.\footnote{For the iDM case, we always assume small enough mass splitting so that the incoming dark sector state can up-scatter if necessary
$E \gg E_{\rm min} = (m_{\chi_2}^2 - m_{\chi_1}^2 + 2 m \, m_{\chi_2})/(2m)$ with $m=m_e, M_A$ so that
 the up-scattering closely follows the standard scattering with a Dirac fermion case~\cite{Batell:2021ooj}.
 }

Given the relatively low kinetic energy of the dark matter particles produced in the late stage of atmospheric showers, we also study the coherent nuclear scattering on a nucleus of mass $M_A$. We follow the treatment of~\cite{Dutta:2019nbn,Dutta:2020vop,CCM:2021leg}  (see also~\cite{Batell:2014yra,deNiverville:2015mwa}) and use:
\begin{align}
    \frac{d \sigma^{cr}}{d E_r} =  Z^2 F_{\textrm{Helm}}^2 (Q)   \frac{ 4 \pi \varepsilon^2 \aem \alpha_D \tilde{f}_{f,s} M_A}{(E^2 - m_\chi^2)(m_{V}^2 + Q^2)}  \ ,
\end{align}
where $Q = \sqrt{2 M_A E_r} $ is the momentum exchanged with the nucleus, $F_{\textrm{Helm}}$ is the Helm form factor and  $\tilde{f}_f(E_r) = (E_r-E)^2+E^2-M_A E_r$  (resp. $\tilde{f}_s(E_r) = (E_r-2E)^2/2$) for a dirac fermion DM (resp. scalar DM). This signal is broadly similar to the recoil from a heavy non-relativist dark matter and can therefore be searched for directly in the various dark matter experiment.

\begin{figure}
	\centering
		\includegraphics[width=0.49\textwidth]{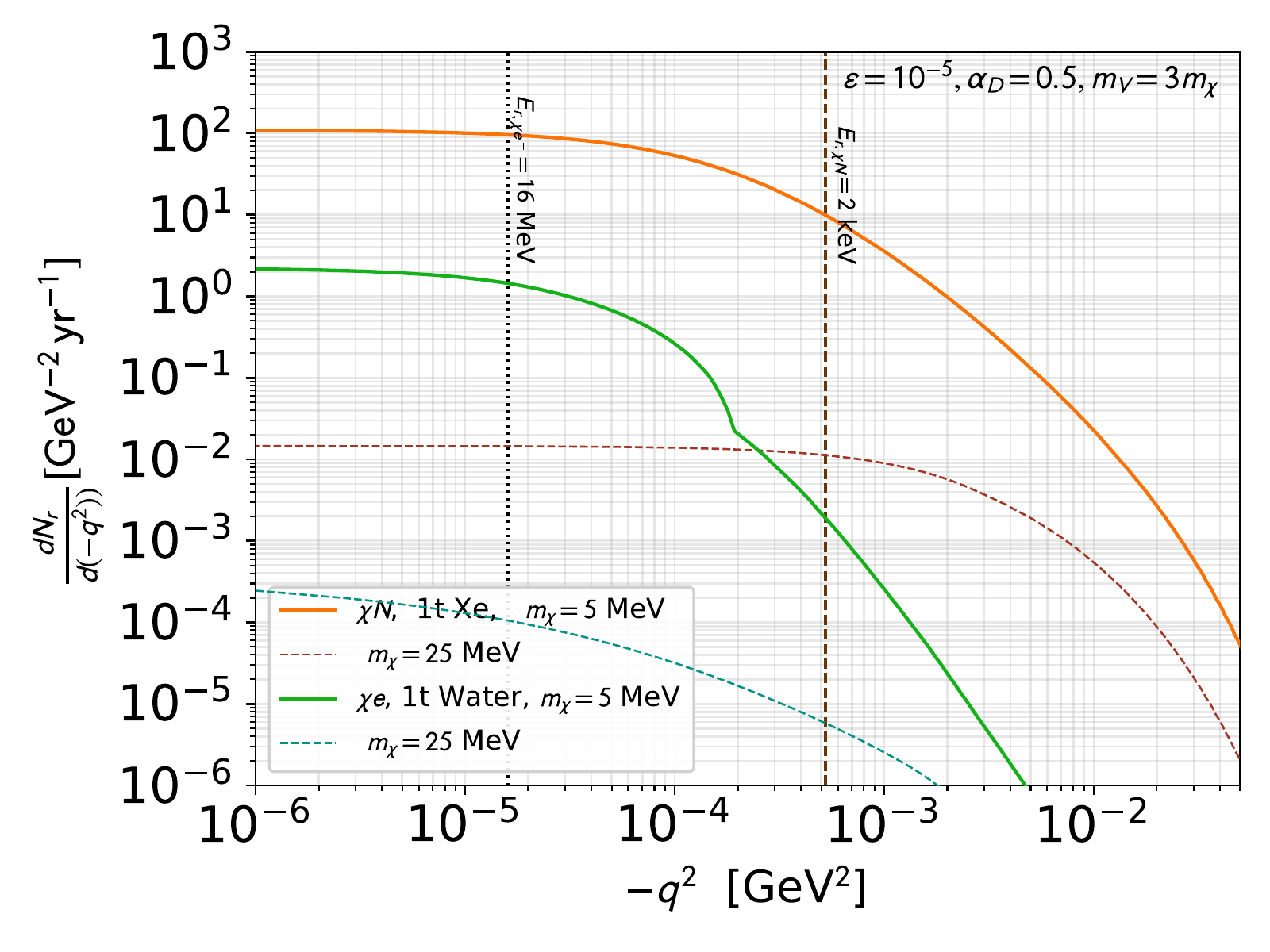}
	\caption{Differential dark matter recoil rates $\partial N_r/\partial(-q^2)$ per $\textrm{ton}\cdot \textrm{yr}$ as function of the squared transferred momentum $-q^2$ for dark matter scattering on electrons in $1$t of water (green lines, $-q^2 ~\equiv~ 2 m_e (E_f-m_e)$) and for coherent scattering on xenon nuclei in $1$t of liquid xenon (orange lines, $-q^2 = 2 m_{\textrm{Xe}} E_r$).  We show the rates for both $m_\chi=5$ MeV and $m_\chi = 25$ MeV. The vertical dotted line corresponds to a electronic recoil of $16$ MeV and the vertical dotted line to a xenon nuclear recoil of $2$ keV. }
			\label{fig:spectrumE}
\end{figure}

We show in Fig.~\ref{fig:spectrumE} the differential number of interactions for a LDM with mass $m_\chi = 5$ MeV and $m_\chi = 25$ MeV as function of the squared exchanged momentum $-q^2$, which is linked to the recoil energy $E_r$ by $-q^2 ~\equiv~ 2 m_{\textrm{Xe}} E_r$ in the coherent scattering case, and by $-q^2 ~\equiv~ 2 m_e (E_f-m_e)$ for electron scattering processes. For the LDM curves with $m_\chi=5$ MeV the rates are dominated by LDM particles from the decay of resonantly-produced dark photon. These dark photons are ``mono-chromatic'' in that they have all the same energy $m_V^2/2m_e$. Thus the dark matter energy spectrum, and consequently the distribution of recoil energies is bounded by this energy, leading in particular to the threshold seen in the $\chi e $ curves in Fig.~\ref{fig:spectrumE}.

\section{Detection in neutrino experiments}
\label{sec:numerics}

\subsection{Experiments considered}

Given that the atmospheric dark matter flux is uniformly spread on earth, the best sensitivity arises from experiments with the largest or most sensitive detection volume. We will illustrate this by considering both the Super-K neutrino telescope program, and the XENON detectors program.

\paragraph{Super-K and upgrades} The super-K detector has an ample dataset of neutrino interactions which can be mimicked by the scattering of a LDM particle. Given that the bulk of the events occurs at low recoil energy, we use the supernova neutrino search from~\cite{Super-Kamiokande:2011lwo}.\footnote{Noting that the spectrum resemble the neutrino relic one in that it is enhanced at low recoil energy.}
The analysis focused on electron recoils between $\sim 16$ MeV to $88$ MeV, and included $2853$ days of data. Following~\cite{Cappiello:2019qsw}, we consider a sensitivity of $~23$ events for the full run and project these results to the HyperK~\cite{Hyper-Kamiokande:2018ofw} design (with a $190$ kt fiducial volume) and the Super-K upgrade with gadolinium doping~\cite{Beacom:2003nk} (SuperK-Gd), assuming an improved sensitivity of $0.84$ (resp. $0.6$ for SuperK-Gd) events per year.

\paragraph{XENON1T and future upgrades} The coherent nuclear scattering signatures discussed previously can be observed in direct detection experiments. We focus on the XENON program, and in particular the data from the XENON1T~\cite{XENON:2018voc} standard WIMP search, which shares most of the characteristic of the coherent scattering signal emphasised above. Including the efficiencies for the selection cuts in the fiducial volume~\cite{XENON:2018voc}, we focus on the intermediate background search corresponding to the $0.9$t reference volume and put a $95$\% C.L limit at $~5$ signal events.\footnote{see Table I of~\cite{XENON:2018voc} and the discussion in the appendix of~\cite{Gondolo:2021fqo}.} We further make projection for $5$ years of data for XENONnT based on the recent projections for the WIMP case~\cite{XENON:2020kmp}: we use a $4$t fiducial volume and estimated nuclear recoil background of $2.0$ events for $20 \textrm{t $\cdot$ yr}$ in  the energy range $4 - 50$ keV. We then scale this result to obtain a projection for five years of data-taking in the DARWIN experiment~\cite{DARWIN:2016hyl} with a $30$t fiducial mass.

\vspace{0.2cm}

We stress that we have focused on these experimental programs primarily as an illustration of the relevance of resonant atmospheric LDM fluxes. Several other experiments could have also sensitivities to this flux. For instance, on the dark matter side, the liquid argon-based DarkSide20k and Argo projects~\cite{DarkSide20k:2020ymr}  or for neutrinos detectors, the DUNE far detector~\cite{DUNE:2020ypp}, KM3-Net~\cite{KM3Net:2016zxf} and JUNO~\cite{JUNO:2015zny} programs.

\subsection{Results}\label{sec:results}

\begin{figure}
	\centering
		\includegraphics[width=0.49\textwidth]{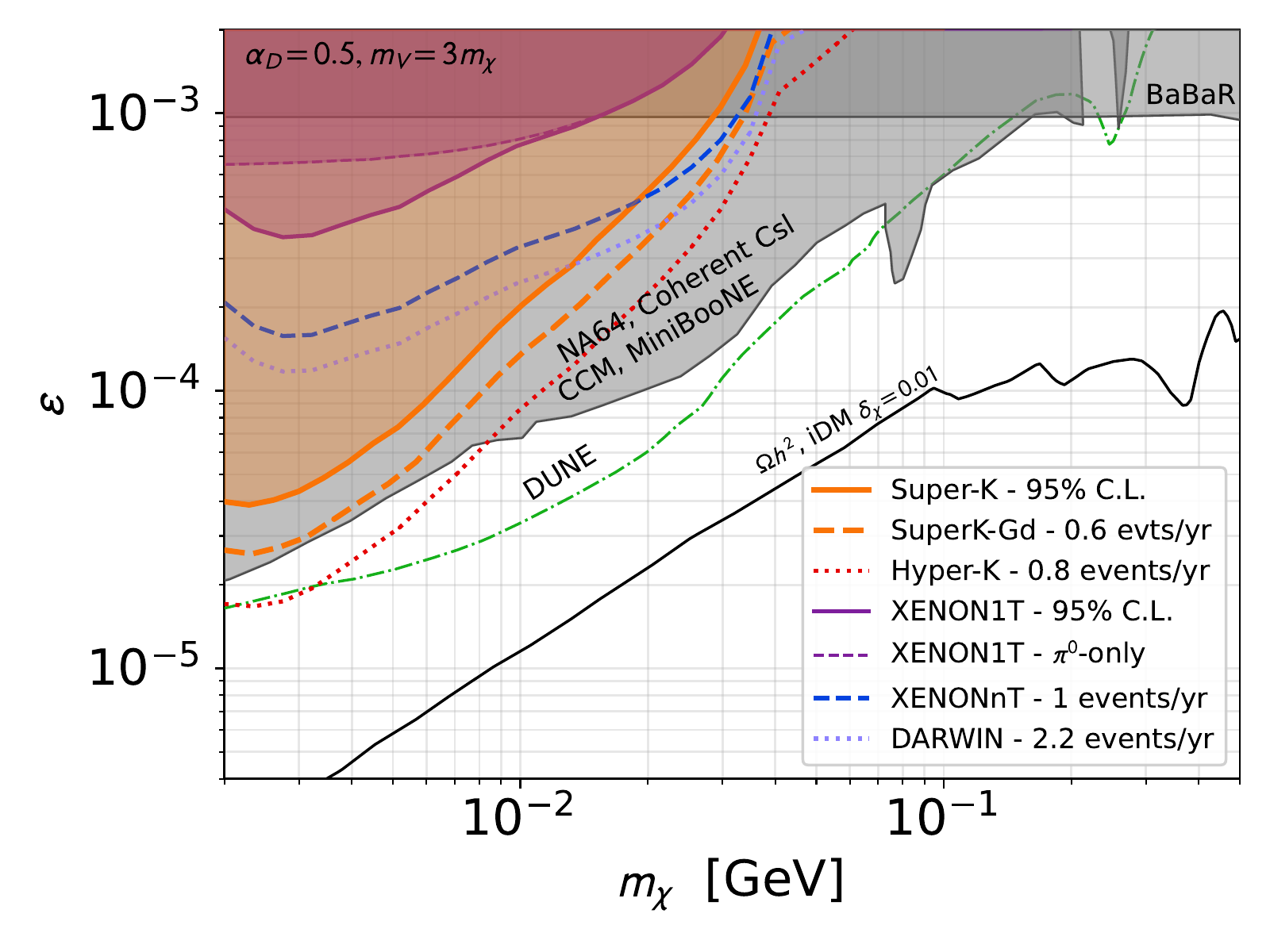}
	\caption{Limit at the 95\% C.L. on our LDM scenario from both the Super-K~\cite{Super-Kamiokande:2011lwo} (orange area) and  XENON1T~\cite{XENON:2018voc} (purple area) experiments, using the ratio $m_V = 3 m_\chi$ and $\alpha_D$. The purple dashed line represent the limit from XENON1T without resonant production. We also show projection for SuperK-Gd with Gadolinium (orange dashed line), and Hyper-K (orange dotted line) following~\cite{Plestid:2020kdm}, and for $5$ years of data in the XENONnT and DARWIN projects~\cite{DARWIN:2016hyl} (blue dashed and dotted lines). The grey regions are the limits from NA64~\cite{Andreev:2021fzd}, MiniBooNE~\cite{MiniBooNEDM:2018cxm}, COHERENT~\cite{COHERENT:2019kwz}, CCM~\cite{CCM:2021leg} and BaBar~\cite{BaBar:2017tiz}. The green dashed-dotted line represents the projection from DM scattering in the near detector of the DUNE experiment as derived in~\cite{Celentano:2020vtu}.}
			\label{fig:limSuperK}
\end{figure}

Resonant production plays a key role in enhancing the production at low masses. We show in Fig.~\ref{fig:limSuperK} the resulting $95 \%$ C.L. limits and projections. The Super-K analysis constrains parameter space comparable to beam neutrino experiments MiniBooNE~\cite{MiniBooNEDM:2018cxm}, COHERENT~\cite{COHERENT:2019kwz} and CCM~\cite{CCM:2021leg} and slightly below the existing limit from the NA64~\cite{Banerjee:2019pds,Andreev:2021fzd}, and BaBar~\cite{BaBar:2017tiz} analysis (see the review~\cite{RF6} for a recent summary). 

The limits for $ 3 m_\chi = m_V \lesssim 40 $ MeV are dominated by the resonant production in the secondary electromagnetic showers, while for   $ 40 \textrm{ MeV} \lesssim m_V $ pion, then $\eta$ meson decays become the dominant production mode.
The sensitivity of nuclear-recoil based analysis in XENON is about an order of magnitude lower than the electron-based in kilotons neutrinos detectors. However, when considering only mesons decays as the production mechanism, it relies only on the LDM interactions with hadronic states and can therefore be used to constrain models where the vector mediator may be lepton or electron-phobic.

We further present projections for the successor of these experiments, starting from the SuperK-Gd experiment with a two-year run with water enriched with Gadolinium~\cite{Beacom:2003nk} and the Hyper-K project, including a long 5-year run and following the background level considered in~\cite{Plestid:2020kdm}. Both improvements push the parameter space in $\varepsilon$ accessible by around a factor of two each. Regarding the case of coherent scattering signal, we also show the projection for both the XENONnT~\cite{XENON:2020kmp} and DARWIN~\cite{DARWIN:2016hyl} (assuming a 5-year run) projects.

The above results can be readily recasted into different types of vector mediators. A particularly relevant case is a baryon-number gauge boson, which presents a natural lepton-phobia. The XENON1T limit $\varepsilon_{X1t}$ based on meson decays only can for instance be projected into a limit on the baryon gauge coupling $g_D$ as $$g_{D, X1t} = \varepsilon_{X1t} \times \sqrt{4 \pi \alpha_{\textrm{em}} Z_{\textrm{Xe}} /  A_{\textrm{Xe}}}$$ comparable to the recent result from the CCM~\cite{CCM:2021leg} collaboration.

\section{Summary}\label{sec:summary}

We have presented in this letter a new atmospheric production mechanism for light dark sector states based on the annihilation of positrons from CR showers on atmospheric electrons. Comparing with the full production from mesons decays (including secondary mesons) that we have obtained from the complete simulation of the CR showers, resonant production dominates at small masses by more than an order of magnitude. It provides an abundant, albeit low energetic, dark matter flux which can be subsequently searched for in detectors with low recoil thresholds. The SM-only distributions as derived from our full numerical simulation are available on a Zenodo database.\footnote{https://zenodo.org/record/6561236} They can be used to estimate the production rates of a large range of bosonic Feebly Interacting Particles, from ALPs to millicharge particles, based on the abundant flux of low energy electrons/positrons and photons generated in the showers. 

We have updated the present and projected constraints from various neutrinos telescope experiments on this class of new physics candidates, focusing in particular on the dark photon-mediated light dark matter scenario. 
The future reach of next generation neutrino telescope is remarkable and on par with accelerator-based experiments, providing a strong incentive for the experimental collaborations to consider this type of analysis in the future. Coherent scattering in next generation dark matter experiments was found to provide weaker limits, but can be used to probe models where the vector mediator may exhibit electron-phobia.

\bigskip
\medskip
\noindent \textbf{Acknowledgments}
\medskip

L.D. thanks E. Nardi, A. Deandrea and the MANOIR group at IP2I for useful discussions, as well as Julien Masbou for details on the XENON program. This work has received partial support by the INFN Iniziativa Specifica Theoretical Astroparticle Physics (TAsP). This project has received funding from the European Union’s Horizon 2020 research and innovation programme under the Marie Sklodowska-Curie grant agreement No. 101028626.

\newpage

\bibliographystyle{utphys}
\bibliography{biblio.bib}

\end{document}